\documentclass[12pt]{iopart}
\usepackage{graphicx}

\begin{document}

\title[Strongly localised molecular orbitals]{Strongly localised molecular orbitals for $\alpha$-quartz}

\author{Oleh Danyliv
\footnote[3]{On leave from Institute for Condensed Matter Physics,
National Academy of Science of Ukraine, Ukraine (e-mail:
oleh.danyliv@kcl.ac.uk)} and Lev Kantorovich}

\address{Department of Physics, Kings College London, Strand, London WC2R
2LS, UK}

\begin{abstract}
A previously proposed computational procedure for constructing a
set of nonorthogonal strongly localised one-electron molecular
orbitals (O. Danyliv, L. Kantorovich - Phys. Rev. B, 2004, to be
published) is applied to a perfect $\alpha$-quartz crystal
characterised by an intermediate type of chemical bonding. The
orbitals are constructed by applying various localisation methods
to canonical Hartree-Fock orbitals calculated for a succession of
finite molecular clusters of increased size with appropriate
boundary conditions. The calculated orbitals span the same
occupied Fock space as the canonical HF solutions, but have an
advantage of reflecting the true chemical nature of the bonding in
the system. The applicability of several localisation techniques
as well as of a number of possible choices of localisation regions
(structure elements) are discussed for this system in detail.
\end{abstract}


\submitto{\JPCM}

\maketitle

\section{Introduction}

A quantum cluster embedding
\cite{Sauer-Sierka-2000,Murphy-Philipp-Freisner-2000,EMC-1,EMC-2,Vreven-Morokuma-2000,Abarenkov-Bulatov-1997}
has become a powerful computational tool in electronic structure
theory of extended systems, such as large biological molecules
\cite{QM/MM,Sauer-Sierka-2000,Hall-Hinde-Burton-Hillier-2000,Rivail3,Murphy-Philipp-Freisner-2000},
surface defects and adsorption on crystal surfaces
\cite{Bredow-1999,Petja-2000,Nasluzov2001} or points defects in
the bulk of crystalline
\cite{Barandiaran-1996,Erbetta-Ricci-Pacchionia-2000} or amorphous
\cite{Sulimov2002} systems. The embedding methods originate from a
model in which a single local perturbation is considered in the
direct space of the entire system inside of a \emph{finite}
quantum molecular cluster in great detail, whereas a more
approximate method is used to account for the rest of the system
surrounding the cluster.

A rather general embedding method is presently being developed in
our laboratory. The central idea of our method is based on the
exact partitioning of the entire system electron density into two
components, one localised within the cluster and the other -
outside it, i.e in the environment. Construction of overlapping
(not orthogonal) strongly localised molecular orbitals (LMO's) as
building blocks of the entire system is crucial for this
technique. The LMO's are designed to represent the true electronic
density of reference systems (such as e.g. 3D ideal perfect
crystals or 2D periodic crystal surfaces) and are constructed to
have transparent chemical meaning, e.g. to represent ions in the
case of ionic systems and covalent bonds in covalently bound
systems. Although we are not yet concerned in this study with
biological systems which do not possess periodicity, we note that
most of the ideas of our method can also be applied to these
systems as well.

A convenient and simple method for calculating the LMO's for 3D
periodic systems (e.g. perfect crystals) was recently suggested in
\cite{Danyliv-LK-2004}. This method is based on finding
appropriate linear combinations of the canonical Hartree-Fock (HF)
solutions for a sequence of finite molecular clusters of increased
size. The linear combinations are chosen in such a way as to
optimise special \emph{localising functionals} constructed to
obtain orbitals localised within certain \emph{regions} (e.g.
bonds, atoms, ions, etc.). There may be several different regions
in every unit cell of the crystal.

The method was successfully applied in \cite{Danyliv-LK-2004} to
two cases of extreme ionic (MgO crystal) and covalent (Si crystal)
bonding. In both cases four LMO's were found in every unit cell.
In the former case every unit cell is composed of a single region
which is associated with an oxygen ion; the region contains eight
electrons and is described by four mutually orthogonal LMO's. In
the latter case (the Si crystal) every unit cell is represented by
four neighbouring regions. Each region is associated with a pair
of nearest Si atoms, contains 2 electrons and is described by a
single double occupied LMO. The four regions belonging to the same
unit cell have one common Si atom at the centre of a tetrahedron
and other four Si atoms form its vertices. The four LMO's within
the same cell do overlap and thus are not orthogonal.

The main purpose of this paper is to check if our method
\cite{Danyliv-LK-2004} can be extended to systems which have more
complicated types of chemical bonding. This is invaluable for the
future development of our embedding method towards describing
insulating crystals with arbitrary type of bonding. Therefore, we
here consider in detail the $\alpha$-quartz (SiO$_{2}$) crystal,
which may be thought of as a prototype system with an intermediate
(ionic-covalent) type of chemical bonding. Essentially two main
questions are addressed here with respect to the localisation of
the calculated LMO's: (i) the choice of regions and (ii) the
choice of localisation methods (i.e. the localisation
functionals).

The plan of the paper is the following. In the next section the
main ideas of our method are briefly described (for the full
discussion, see \cite{Danyliv-LK-2004}). Three localisation
methods are introduced (one of them was not used in our previous
study \cite{Danyliv-LK-2004}) alongside with a choice of three
localisation criteria. Application of our method to the
$\alpha$-quartz crystal is considered in Section
\ref{sec:SiO2-bulk}. Brief conclusions are given in Section
\ref{sec:Discussion-and-conclusions}.

\section{Localisation procedure \label{sec:Localisation-methods}}

\subsection{General approach}

Let us assume that we know an occupied canonical set of
one-electron molecular orbitals, $\left\{
\varphi_{i}^{c}(\mathbf{r)}\right\} $, for a perfect 3D periodic
crystal. These orbitals may be obtained as eigenvectors of the
appropriate Hartree-Fock (HF) problem using e.g. the CRYSTAL code
\cite{CRYSTAL98} which employs directly the periodic symmetry. In
our method, however, we consider instead a specially designed set
of \emph{finite} clusters of increased size and find the HF
solutions for them using one of the available quantum chemistry
packages. It was demonstrated in \cite{Danyliv-LK-2004} that this
approach is equivalent to using a periodic-crystal electronic
structure approach as far as the LMO's are concerned, provided
that large enough molecular clusters are used.

The canonical molecular orbitals (CMO's) are orthogonal and span
the entire occupied Fock space. They are not localised in space,
and have a non-zero contribution on atoms of every unit cell in
the crystal. In practice, when the cluster method is employed,
they span the entire cluster. In other words, the CMO's are
assumed to be given as a linear combination of the atomic
orbitals, $\chi_{\mu}(\mathbf{r})$, centred on all atoms of the
cluster in question: \begin{equation}
\varphi_{i}^{c}(\mathbf{r})=\sum_{\mu}C_{\mu
i}^{c}\chi_{\mu}(\mathbf{r})\label{eq:CO_via_AO}\end{equation}

In order to describe the crystal as a set of overlapping
(non-orthogonal) localised functions, $\left\{
\widetilde{\varphi}_{a}(\mathbf{r)}\right\} $, which span the same
occupied Fock space, one has to obtain appropriate linear
combinations of the original canonical set $\left\{
\varphi_{i}^{c}(\mathbf{r)}\right\} $. In order to do this, it is
first necessary to identify \emph{regions} of space where each of
the functions $\widetilde{\varphi}_{a}(\mathbf{r)}$ has to be
localised. Although any (non-singular) linear combination of the
canonical set will give the same electron density
$\rho(\mathbf{r})$, we adopt here a strategy based on the
chemistry of the system in question. Namely, the choice of the
localisation regions in the first instance is based on the
expected type of the chemical bonding in the system, e.g.
atoms/ions in the cases of atomic/ionic systems, two nearest atoms
in the case of covalent bonding, etc. A more complicated choice is
anticipated in the cases of intermediate bonding as will be
demonstrated in Section \ref{sec:SiO2-bulk}. Several different
nonequivalent regions may be necessary to represent a crystal unit
cell which can then be periodically translated to reproduce the
whole infinite crystal. Note that several localised orbitals may
be associated with each region. For instance, in the case of the
Si crystal one needs four localised regions associated with four
bonds; each bond is represented by a single double occupied
localised orbital and all four bonds have one common Si atom in
the centre of the tetrahedron.

Once the localised regions are identified, it is necessary to find
linear combination of the CMO's which are localised in each of the
regions,\begin{equation}
\widetilde{\varphi}_{a}(\mathbf{r})=\sum_{j}^{occ}U_{aj}\varphi_{j}^{c}(\mathbf{r})\equiv\sum_{\mu}\widetilde{C}_{\mu
a}\chi_{\mu}(\mathbf{r})\label{eq:U-transf}\end{equation}
 The transformation $\mathbf{U=\parallel}U_{aj}\mathbf{\parallel}$
of the CMO's within the occupied subspace is arbitrary and, in
general, \emph{non-unitary}. In the latter case the expression for
the density via the new set of orbitals should contain the inverse
of the overlap matrix
$\mathbf{\widetilde{S}=\parallel}\widetilde{S}_{ab}\mathbf{\parallel}$
\cite{McWeeny}: \begin{equation}
\rho(\mathbf{r})=2\sum_{ab}^{occ}\widetilde{\varphi}_{a}(\mathbf{r})\left(\widetilde{\mathbf{S}}\right)_{ab}^{-1}\widetilde{\varphi}_{b}^{*}(\mathbf{r})\label{eq:density-non-orth}\end{equation}
where $\widetilde{S}_{ab}=\left\langle
\widetilde{\varphi}_{a}(\mathbf{r})\right|\left.\widetilde{\varphi}_{b}(\mathbf{r})\right\rangle
$ is the overlap integral. The double summation here is performed
over all localised orbitals of the whole infinite crystal. If the
transformation is unitary, then both the overlap matrix and its
inverse are unity matrices and the density takes on its usual
{}``diagonal{}`` form.

In general, any localisation procedure is equivalent to some
transformation $\mathbf{U}$ of the CMO's. To find the necessary
transformation for, say, region $A$, an optimisation (minimisation
or maximisation) problem is formulated for some specific
\emph{localising functional} $\widetilde{\Omega}_{A}\left[\left\{
\widetilde{\varphi}_{a}\right\} \right]$ with the constraint that
the LMO's associated with region $A$ are orthonormal. This leads
to a standard eigenvalue-eigenvector problem:
\begin{equation}
\sum_{j}^{occ}\Omega_{ij}^{A}U_{aj}=\lambda_{a}U_{ai}\label{eq:eigenv-problem-matrix}\end{equation}
for the elements of the transformation matrix $\mathbf{U}$. Here
$\Omega_{ij}^{A}$ is a matrix element of an operator
$\widehat{\Omega}_{A}$ calculated using canonical orbitals
$\varphi_{i}^{c}(\mathbf{r)}$. The operator $\widehat{\Omega}_{A}$
is uniquely defined from the functional
$\widetilde{\Omega}_{A}\left[\left\{
\widetilde{\varphi}_{a}\right\} \right]$. Although for some
localising functionals (see, e.g. \cite{Danyliv-LK-2004}) the
matrix elements $\Omega_{ij}^{A}$ may depend on the LMO's
themselves so that the problem (\ref{eq:eigenv-problem-matrix}) is
to be solved self-consistently, we do not consider those
functionals in this paper.

Note, that LMO's associated with different regions will not be
orthogonal in this method. This is because they are obtained from
different localising functionals which strongly depend on the
region in question, so that LMO's from different regions are
determined by solving different secular problems. For instance, if
LMO's $\left\{ \widetilde{\varphi}_{a}(\mathbf{r)}\right\} $
correspond to region $A$, then the LMO's $\left\{
\widetilde{\varphi}_{b}(\mathbf{r)}\equiv\widetilde{\varphi}_{a}(\mathbf{r-L}\mathbf{)}\right\}
$ are obtained for a physically equivalent region $B$ separated
from $A$ by a lattice translation $\mathbf{L}$.

Using a physical argument, each region is associated with a
certain \emph{even} number of electrons $2n$. Therefore, if
$\Omega_{A}$ is minimised, we choose the first $n$ eigenvectors of
the problem (\ref{eq:eigenv-problem-matrix}); if, however,
$\Omega_{A}$ is maximised, the last $n$ solutions are adopted. By
collecting LMO's from all regions in the unit cell and then
translating those over the whole crystal it should be possible to
span the whole occupied Fock space and thus construct the total
electron density (\ref{eq:density-non-orth}). The larger the
finite cluster used while calculating the canonical orbitals, the
closer the Fock space will be reproduced by the LMO's.

To summarise, we first suggest a possible set of localisation
regions in the unit cell and then consider a set of finite
molecular clusters (with appropriate boundary conditions) which
have all these regions in their central part. Then, we obtain the
occupied canonical HF orbitals for each of the clusters using a
standard quantum-chemistry package. Out of all the clusters
considered, a cluster is chosen for which the electron density is
well converged in its central part. Next, using a localisation
functional, canonical occupied HF orbitals of the chosen cluster
are transformed into LMO's. The procedure is repeated for several
localisation functionals and in each case localisation criteria
are applied. Then, if necessary, a different choice of
localisation regions is made, and the whole procedure is repeated.
As will be seen in Section \ref{sec:SiO2-bulk}, in the case of the
SiO$_{2}$ crystal, three different sets of regions can be
suggested; however, the same set of clusters will be used to
calculate the LMO's in each case.

\subsection{Localising functionals}

In a number of methods \cite{Mayer1996} the localising functionals
are proportional to the non-diagonal electron {}``density''
associated with region A, \begin{equation}
\sigma_{A}(\mathbf{r},\mathbf{r}^{\prime})=\sum_{a=1}^{n}\widetilde{\varphi}_{a}(\mathbf{r})\widetilde{\varphi}_{a}^{*}(\mathbf{r}^{\prime})\label{eq:region-A_density}\end{equation}
where the summation is performed over all $n$ LMO's of region $A$.
Note that for convenience of the final equations we have omitted a
factor of two above as it is unimportant for the eigenproblem
(\ref{eq:eigenv-problem-matrix}) to be solved. Therefore, the
functionals can be represented in the following general
form:\begin{equation}
\Omega_{A}=\int\left[\widehat{\Omega}_{A}\sigma_{A}(\mathbf{r},\mathbf{r}^{\prime})\right]_{\mathbf{r^{\prime}\rightarrow
r}}d\mathbf{r}=\sum_{a=1}^{n}\int\widetilde{\varphi}_{a}^{*}(\mathbf{r})\widehat{\Omega}_{A}\widetilde{\varphi}_{a}(\mathbf{r})d\mathbf{r}\equiv\sum_{a=1}^{n}\sum_{jk}^{occ}U_{aj}^{*}\Omega_{jk}^{A}U_{ak}\label{eq:F_A_first_group}\end{equation}
where $\widehat{\Omega}_{A}$ is the localisation operator and the
Hermitian matrix
$\mathbf{\Omega}^{A}=\parallel\Omega_{jk}^{A}\parallel$ can easily
be written in terms of the canonical MO's using the definition
(\ref{eq:CO_via_AO}):

\begin{equation}
\Omega_{jk}^{A}=\left\langle
\varphi_{j}^{c}\right|\widehat{\Omega}_{A}\left|\varphi_{k}^{c}\right\rangle
=\sum_{\mu,\nu}C_{\mu j}^{c*}C_{\nu k}^{c}\left\langle
\chi_{\mu}\right|\widehat{\Omega}_{A}\left|\chi_{\nu}\right\rangle
\label{eq:Q-matrix-1st-group}\end{equation} For all methods to be
considered below both the operator $\widehat{\Omega}_{A}$ and the
matrix $\mathbf{\Omega}^{A}$ do not depend on the LMO's sought
for, so that in order to obtain the localised orbitals one has
simply to find the eigenvectors of the matrix
$\mathbf{\Omega}^{A}$ using Eq. (\ref{eq:eigenv-problem-matrix}).
Three particular localisation methods implemented in this work are
considered in the following in more detail. Note that one of the
methods (method G) was not considered in \cite{Danyliv-LK-2004}.

\subsubsection{Mulliken's net population (method M)}

In this method the localisation region A is specified by a
selection of AO's (e.g. on one or two particular atoms in the unit
cell). Then, the net atomic Mulliken's \cite{Mulliken} population
produced by the LMO's in the selected region is maximised
\cite{Magnasco-Perico,Mayer1996,Danyliv-LK-2004}. In this case
\begin{equation} \Omega_{jk}^{A}=\sum_{\mu,\nu\in A}C_{\mu
j}^{c*}S_{\mu\nu}C_{\nu
k}^{c}\label{eq:Q-matrix_for_Mulliken}\end{equation} where
$S_{\mu\nu}$ is the overlap integral between two AO's $\chi_{\mu}$
and $\chi_{\nu}$. The summation here is performed over AO's which
are centred in the chosen region $A$. This way one can make the
LMO's to have a maximum contribution from the specified AO's in
region $A$. Sometimes a different choice of AO's centred on the
\emph{same} atoms may lead to physically identical localisation;
however, this is not the case in general \cite{Danyliv-LK-2004}.
This method will be referred to as method M.

\subsubsection{Mulliken's gross population (method G)}

If, instead, the Mulliken's gross population on the atoms
belonging to region A is maximised, one arrives into the
Pipek-Mezey localisation scheme \cite{Pipek-Mezey,Mayer1996}. In
this case the expression for $\Omega_{jk}^{A}$ is very similar to
that given by Eq.
(\ref{eq:Q-matrix_for_Mulliken}):\begin{equation}
\Omega_{jk}^{A}=\frac{1}{2}\sum_{\mu\in A}\sum_{\nu}\left\{ C_{\mu
j}^{c*}S_{\mu\nu}C_{\nu k}^{c}+C_{\nu j}^{c*}S_{\nu\mu}C_{\mu
k}^{c}\right\} \label{eq:Q-matrix_for_Pipek-Mezey}\end{equation}
The first summation here is performed over AO's which are centred
in the chosen region $A$ and another summation is performed over
all AO's. This method will be referred to as method G.

\subsubsection{The projection on the atomic subspace (method P)}

The Roby's population maximisation \cite{Roby74} gives LMO's for
which the projection on the subspace spanned by the basis orbitals
centred within the selected region A is a maximum, or is at least
stationary \cite{Mayer1996,Danyliv-LK-2004}. In this method the
localisation operator $\widehat{\Omega}_{A}$ in Eq.
(\ref{eq:F_A_first_group}) is chosen in the form of a projection
operator, so that: \begin{equation}
\Omega_{jk}^{A}=\sum_{\lambda,\tau}C_{\lambda j}^{c*}C_{\tau
k}^{c}\left[\sum_{\mu,\nu\in
A}S_{\lambda\mu}(\mathbf{S}_{A}^{-1})_{\mu\nu}S_{\nu\tau}\right]\label{eq:Q-matrix_Oper_method}\end{equation}
where $\mathbf{S}_{A}^{-1}$ stands for the inverse of the overlap
matrix $\mathbf{S}_{A}$ defined on all AO's $\mu,\nu\in A$. Here
the first double summation is performed over all AO's of the
system. Note that the idempotent operator $\widehat{\Omega}_{A}$
projects any orbital into a subspace spanned by the AO's
associated with region $A$ only. Therefore, by choosing particular
AO's (and thus the region) one ensures the maximum overlap of the
LMO's with them. It is seen that this method, which will be
referred to as method P, although different in implementation, is
very similar in spirit to the previous two methods.

\subsection{Localisation criteria\label{sub:Localization-criteria}}

An application of the various schemes described above results in
LMO's which are localised in the 3D space differently. It is
therefore useful to have simple criteria which can identify the
degree of their localisation. Note that each of the localisation
methods of the previous subsection corresponds to a particular
linear combination of the canonical orbitals and thus will result
in exactly the same electron density (\ref{eq:density-non-orth})
provided, of course, that a sufficiently large cluster has been
used in the LMO's calculation. We assume in this section that this
is always the case.

Three methods will be used to assess the localisation of
calculated LMO's \cite{Danyliv-LK-2004}.

\subsubsection{Localisation index}

The first method was proposed by Pipek and Mezey
\cite{Pipek-Mezey} and is based on the calculation of the
so-called \textit{localisation index}\textit{\emph{:}}

\begin{equation}
d_{a}=\left\{ \sum_{B}\left[\sum_{\mu\in
B}\sum_{\nu}\left(\widetilde{C}_{\mu
a}S_{\mu\nu}\widetilde{C}_{\nu a}\right)\right]\right\}
^{-1}\label{eq:Loclization index}\end{equation} where the first
summation is run over all atoms B of the entire system. Here the
quantity in the square brackets is similar to the diagonal part of
the localisation operator matrix
(\ref{eq:Q-matrix_for_Pipek-Mezey}) calculated on real localised
orbitals. Qualitatively, the localisation index gives the number
of atoms on which the orbital
$\widetilde{\varphi}_{a}(\mathbf{r})$ is predominantly localised.
Therefore for the ionic type of bonding one would expect
$d_{a}\sim1$ and for a valence LMO describing a covalent bonding
$d_{a}\sim2$.

\subsubsection{Eigenvalues of the overlap matrix}

Alternatively, the overlap between LMO's also gives an important
information about their localisation. That is why as the second
criterion we shall consider the maximum eigenvalue of the overlap
matrix. Note that for periodic structures it is more convenient to
use the Fourier transformation of the overlap matrix
\cite{Danyliv-LK-periodic-2004}:

\begin{equation}
S_{ab}(\mathbf{k})=\sum_{\mathbf{L}}\left\langle
\widetilde{\varphi}_{a}(\mathbf{r})\left|\widetilde{\varphi}_{b}(\mathbf{r-L})\right.\right\rangle
\mathrm{e}^{i\mathbf{kL}}\label{eq:Q-matrix_for_critirea}\end{equation}
where $\mathbf{k}$ is a point in the Brillouin zone,
$\widetilde{\varphi}_{a}(\mathbf{r})$ and
$\widetilde{\varphi}_{b}(\mathbf{r})$ are LMO's in the zero
(central) elementary unit cell and $\mathbf{L}$ is the lattice
translation vector. Note, that if any of the eigenvalues,
$\lambda(\mathbf{k)}$, of the overlap matrix
$\mathbf{S}=\left\Vert S_{ab}(\mathbf{k})\right\Vert $ is found to
be larger than 2, it is impossible to obtain the total crystal
density in this basis by expanding the inverse of the overlap
matrix in Eq. (\ref{eq:density-non-orth}) in powers of the overlap
(the L\"owdin's method, see \cite{Danyliv-LK-periodic-2004} for a
detailed discussion). Therefore, the existence of large
eigenvalues of the matrix $\mathbf{S}$ corresponds to a weak
localisation of the LMO's.

\subsubsection{Gap in eigenvalues of the localisation problem}

The eigenvalues $\lambda_{a}$ of the secular problem
(\ref{eq:eigenv-problem-matrix}) can also be used to indicate the
degree of localisation \cite{Danyliv-LK-2004}. Indeed, if the
localisation functional $\Omega_{A}$ used is appropriate, then (i)
the chosen $n$ solutions would have close eigenvalues
$\lambda_{a}$ which correspond to their similar localisation in
region $A$, and (ii) the gap $\Delta\lambda$ in the eigenvalues
$\lambda_{a}$ between the chosen $n$ and other solutions is
considerable, i.e. the other solutions will correspond to much
worse localisation in region $A$ (cf. \cite{Whitt-Pakk}).
Therefore, in order to check the localisation of the LMO's, we
shall also use the parameter $\Delta\lambda$.

One may assume that better localisation will give larger gap value
$\Delta\lambda$. In principle, if the given region and the right
number of LMO's $n$, associated with it, were chosen correctly,
one should expect some gap $\Delta\lambda$ in the eigenvalues
$\lambda_{a}$.

\section{SiO$_{2}$ bulk\label{sec:SiO2-bulk}}

\begin{figure}
\begin{center}\includegraphics[%
  height=7cm,
  keepaspectratio]{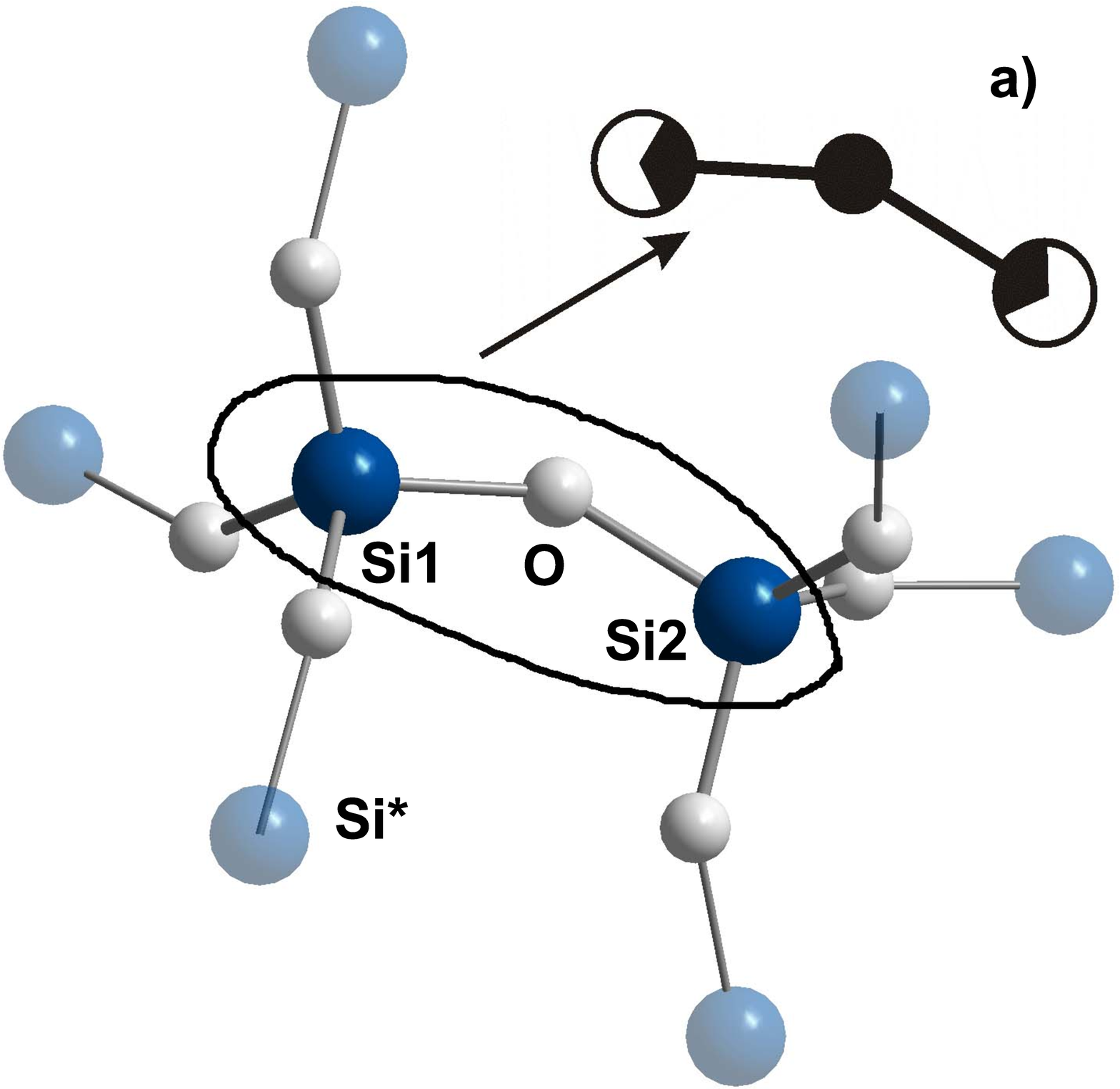}\end{center}

\begin{center}\includegraphics[%
  width=100cm,
  height=7cm,
  keepaspectratio]{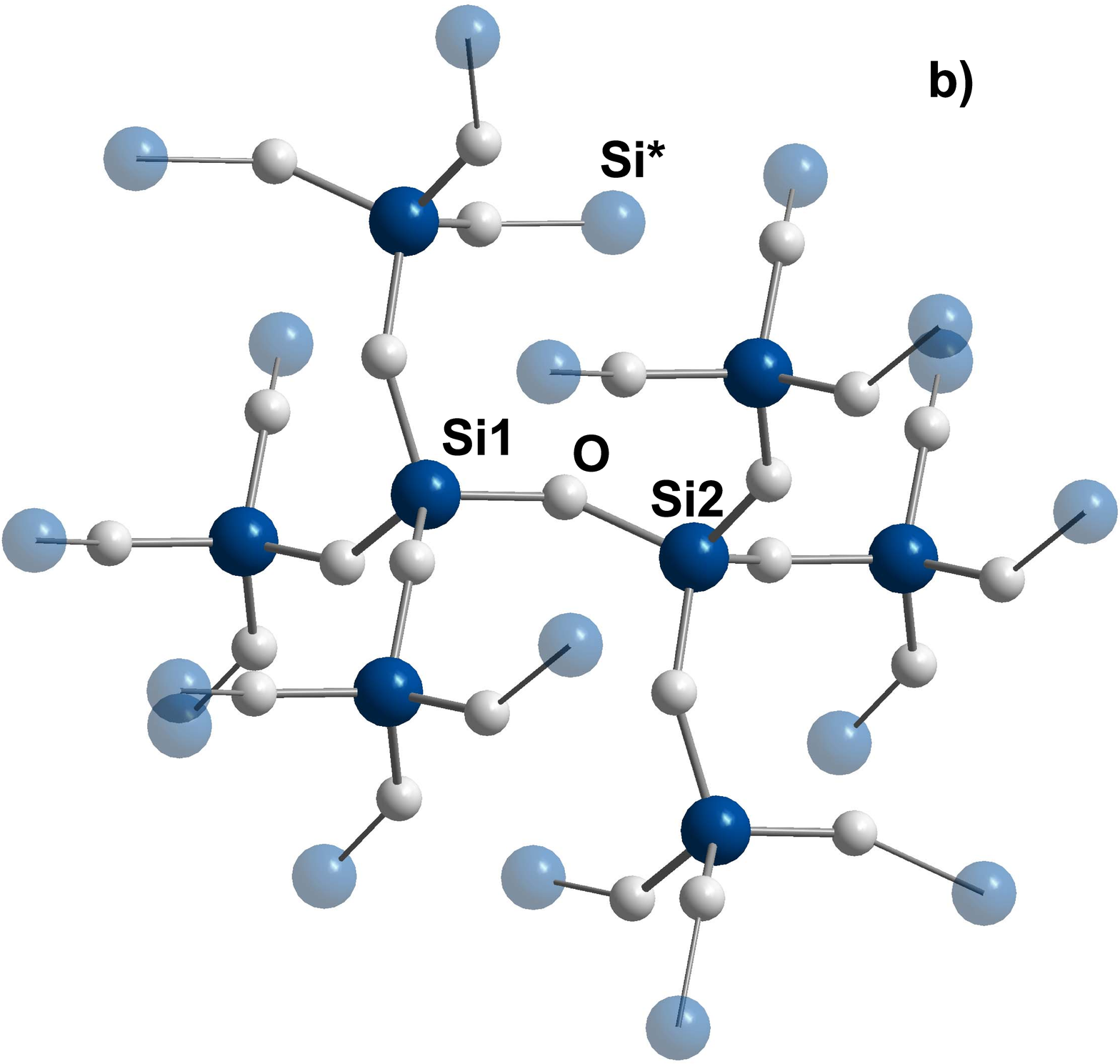}\end{center}

\caption{The first two quantum clusters, Si$_{2}$O$_{7}$ (a) and
Si$_{8}$O$_{25}$ (b), used in our HF calculations. Point charges
surrounding the clusters are not shown. Both silicon atoms (Si1
and Si2) and the oxygen atom of the central unit
Si$_{\frac{1}{4}}$OSi$_{\frac{1}{4}}$ are indicated in each case.
The elementary {}``brick'' the whole system can be composed from
is shown schematically in (a). \label{Fig:size}}
\end{figure}

$\alpha$-quartz (SiO$_{2}$) crystal has a hexagonal Bravais
lattice and corresponds to the $D_{3}^{4}$ ($P3_{1}21$, No. 152)
space group symmetry \cite{Wyckoff,LevKant-book}. The equilibrium
crystal structure has been found using the Density Functional
Theory, plane wave basis set, periodic boundary conditions and the
method of ultrasoft pseudopotentials as implemented in the VASP
code \cite{VASP1,VASP2,Kresse-softPP}. The calculation was
carefully converged with respect to the $\mathbf{k}$ point
sampling and the plane wave cut-off. The lattice found is
specified by elementary translations $\mathbf{a}_{1}=a(0,-1,0)$,
$\mathbf{a}_{2}=a(\frac{1}{2},\frac{\sqrt{3}}{2},0)$ and
$\mathbf{a}_{3}=c(0,0,1)$ with $a$=4.913 \AA$\,$ and $c$=5.4046
\AA. Each cell contains nine atoms distributed over three
SiO$_{2}$ molecules. Three $3a$ positions (the local symmetry
$C_{2}$) are occupied by Si atoms, the position of the first Si
atom is given by the fractional coordinates $(-u,-u,\frac{1}{3})$
with $u$=0.4697; six oxygen atoms occupy a general position $6c$
which can be generated from the fractional coordinates $(x,y,z)$
using $x$=0.4135, $y$=0.2669 and $z$=0.1191. The tetrahedron of
oxygen atoms with a silicon atom in its centre is almost regular
with slightly different Si-O distances of $R$(Si1-O)=1.613 \AA$\,$
and $R$(Si2-O)=1.604 \AA. The whole 3D crystal structure can be
constructed from Si$_{\frac{1}{4}}$OSi$_{\frac{1}{4}}$ units
connected together at the Si atoms as shown schematically in Fig.
\ref{Fig:size}(a). Four such units have a common Si atom at the
centre of a tetrahedron, six complete inequivalent units
(positioned differently in space) form an elementary cell. Note
that this is similar to the Si crystal structure where the whole
crystal can be composed of Si$_{\frac{1}{4}}$Si$_{\frac{1}{4}}$
units \cite{Danyliv-LK-2004}.

Because of such an arrangement of atoms in the $\alpha$-quartz
crystal, it is reasonable to assume that every unit
Si$_{\frac{1}{4}}$OSi$_{\frac{1}{4}}$ forms one independent
region. However, as will be shown in the following, one can
alternatively consider two or three regions made out of each unit
as well. Therefore, in choosing a set of finite clusters, we
ensured that the whole Si$_{\frac{1}{4}}$OSi$_{\frac{1}{4}}$ unit
was in the centre of each of the quantum clusters. Three clusters
were considered: Si$_{2}$O$_{7}$, Si$_{8}$O$_{25}$ and
Si$_{40}$O$_{103}$, containing 9, 33 and 143 \textbf{}atoms,
respectively. The smallest and middle size clusters used are shown
in Fig. \ref{Fig:size}. To create proper termination of the
clusters at their boundary, we implemented special boundary
conditions suggested in \cite{Sulimov2002}. In particular, we
surrounded clusters by pseudoatoms Si{*} which are directly
connected to the cluster O atoms. Each Si{*} pseudoatom is made of
{}``classical (3/4-th) and {}``quantum'' (1/4-th) parts. The
{}``classical'' part is represented by the electrostatic potential
due to a +1.8$e$ point charge (which is a 3/4-th of the effective
charge on a Si atom in the lattice), $e$ being the elementary
charge. The {}``quantum'' part of the Si{*} pseudoatom consists of
a central repulsive electronic potential $V(r)$ (added to mimic
the screening of the Si core by the valence electrons) and a
single valence electron. The parameters of the potential and the
basis set for the pseudoatoms were optimised to get proper
effective charges on Si and O atoms and to eliminate the
contribution of the Si{*} electron at the top of the valence and
the bottom of the conduction bands. Note that the described
boundary conditions were found to be crucial only for the smallest
cluster; for other two clusters simpler boundary conditions (e.g.
termination by hydrogen atoms) were also tried and found to give
practically identical results for the LMO's and thus will not be
discussed further. To simulate the Madelung field, clusters were
surrounded by an array of nearly $2.5\times10^{4}$ point charges,
containing +2.4$e$ charges to mimic Si atoms and -1.2$e$ charges
for oxygens.

To simplify the initial HF calculations required to check the
convergence of the electron density with the cluster size and
generate all occupied canonical molecular orbitals, only valence
(\textit{$3s^{2}3p^{2}$} on Si and $2s^{2}2p^{4}$ on O) electrons
were considered explicitly by using for both species the coreless
HF pseudopotentials (CHF) with LP-31G basis set from Ref.
\cite{CHF:Melius-Goddard74}. The calculations were made with the
use of the Gamess-UK package \cite{GAMESS-UK}.

\begin{figure}
\begin{center}\includegraphics[%
  width=100cm,
  height=7cm,
  keepaspectratio]{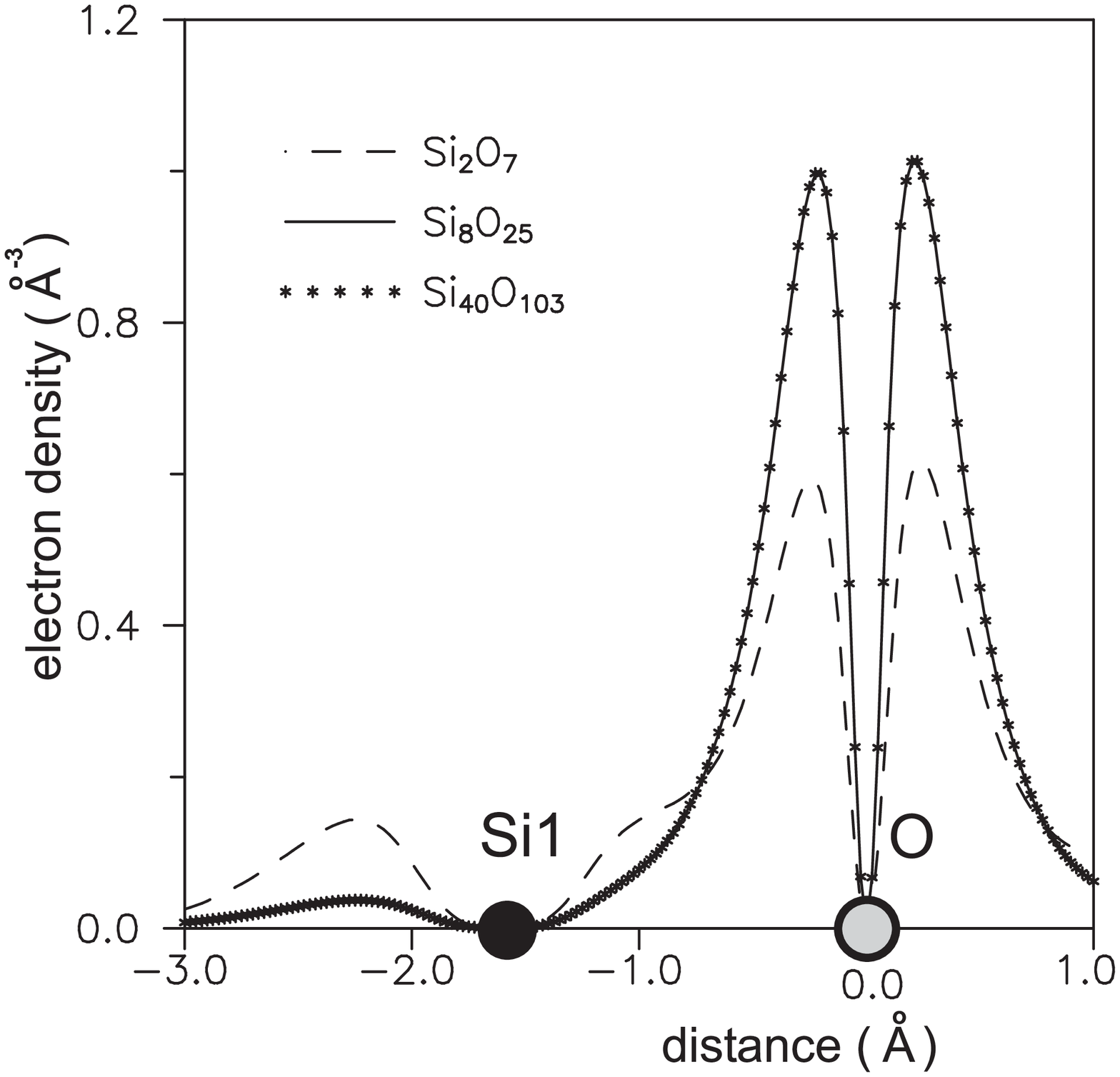}\end{center}

\caption{The HF electron densities for the three clusters are
plotted along the Si1-O direction. The convergence is obvious: the
curves for the middle sized and the largest clusters are
indistinguishable. \label{Fig:convergence}}
\end{figure}

Fig. \ref{Fig:convergence} shows the convergence of the HF
electron density with the size of the cluster along the Si1-O
direction. We carefully checked, by plotting the densities along
other directions and by making 2D plots, that this direction is
representative for assessing the convergence in this system. The
deep minimum in the density at the oxygen atom is due to the
pseudopotential method used. One can see that the difference
between curves for the middle sized and the largest clusters is
negligible. This suggests that the middle sized cluster
Si$_{8}$O$_{25}$ is perfectly sufficient for our purposes
and thus was used in all the calculations described below. %

\begin{figure}
\begin{center}\includegraphics[%
  width=100cm,
  height=7cm,
  keepaspectratio]{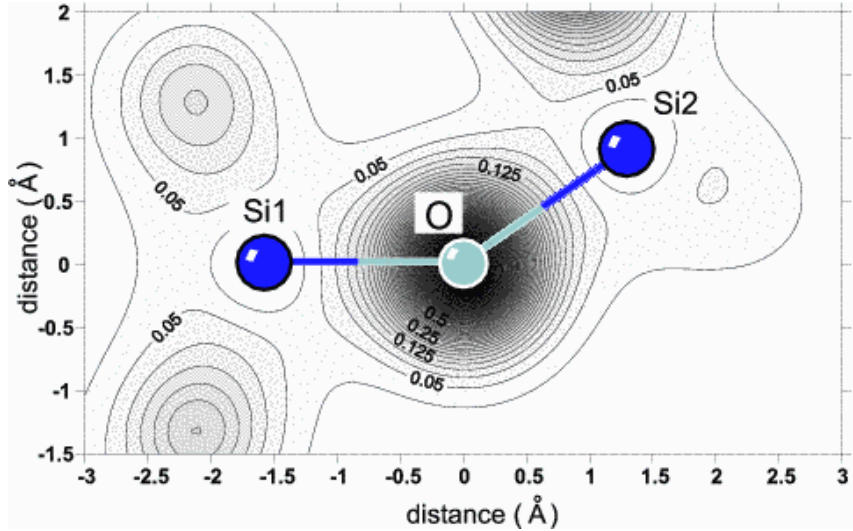}\end{center}

\caption{The total electron valence density $\rho(\mathbf{r})$ in
the Si1-O-Si2 plane by combining the contributions of LMO's
(method M, model 1) across the infinite crystal. The density was
calculated using the method \cite{Danyliv-LK-periodic-2004} based
on Eq. (\ref{eq:density-non-orth}). \label{Fig:density2D} }
\end{figure}

As has been mentioned above, the elementary {}``brick'' we can
build the system from is the unit
Si$_{\frac{1}{4}}$OSi$_{\frac{1}{4}}$ shown in the centre of each
of the clusters in Fig.\ref{Fig:size} as a Si1-O-Si2 molecule.
Each such unit should be assigned 8 electrons in total: 6
electrons come from the O atom and by 1 electron from each of the
two Si atoms. Note that each Si atom contributes to four different
units which contain this Si atom. This simple analysis allows us
to suggest at least three possible choices (models) for the
localisation regions:

\begin{enumerate}
\item Each unit Si$_{\frac{1}{4}}$OSi$_{\frac{1}{4}}$containing 8
electrons is considered as a \emph{single} region; therefore, to
obtain the corresponding \emph{four} (double occupied) LMO's, one
has to choose all AO's centred on the atoms Si1, Si2 and O in the
centre of the cluster when applying any of the localising
functionals discussed above. Note that all four LMO's obtained
using this partition method will be orthonormal as eigenvectors of
the same secular problem (\ref{eq:eigenv-problem-matrix}). \item
Each pair of atoms Si1-O and Si2-O can be considered as a separate
region, i.e. there will be \emph{two} regions in total to describe
every unit Si$_{\frac{1}{4}}$OSi$_{\frac{1}{4}}$; four electrons
distributed over two (double occupied) LMO's will be associated in
this case with each of the two regions. Two LMO's associated with
either of the two regions will be mutually orthogonal; however,
the LMO's belonging to different regions will have a non-zero
overlap. \textbf{} \item Finally, each unit can be split into
\emph{three} different regions: (i) the first region, containing
two electrons and described by a single LMO, is constructed to
describe a covalent bond Si1-O; this can be achieved by enforcing
localisation on $2p$ AO's of the O atom and all AO's of the Si1
atom; (ii) the second region is formed similarly to describe the
Si2-O bond; (iii) finally, the remaining four electrons are
attached to the third region which is localised predominantly on
the O atom giving rise to two more (double occupied) LMO's; in
this case the $2s$ AO's centred on the O atom can be used to
inforce localisation. Thus, in this case there will be three sets
of the LMO's: two orthogonal LMO's belonging to the O region and
other two LMO's belonging to the {}``bond'' regions; the latter
two LMO's have a non-zero overlap with any other LMO.
\end{enumerate}
For each choice of the localisation regions described above (which
will be referred to as \emph{regions models} hereafter), we can
apply either of the three localisation methods of Section
\ref{sec:Localisation-methods} to calculate the LMO's using the
obtained occupied canonical orbitals of the middle cluster. When
solving the secular problem (\ref{eq:eigenv-problem-matrix}), the
contributions of the boundary pseudoatoms Si{*} were removed from
the canonical orbitals which then were renormalised. Using the
obtained LMO's, one should appropriately translate and rotate them
in order to obtain all LMO's comprising the whole primitive cell.
(For instance, there will be six sets with four LMO's in each in
the primitive cell for model 1.) By applying lattice translations
to all the LMO's associated with the primitive cell, the whole
infinite crystal is reproduced. It is then possible to calculate
the total electron density $\rho(\mathbf{r)}$ of the whole crystal
using Eq. (\ref{eq:density-non-orth}). The necessary lattice
summations are handled exactly by converting into the $\mathbf{k}$
space \cite{Danyliv-LK-periodic-2004}. These calculations have
been done for all nine cases (three localisation methods versus
three choices of the localisation regions). The calculated
$\rho(\mathbf{r)}$ matched exactly the original HF density in the
central part of the cluster in all cases indicating that a very
good degree of localisation was achieved in each case. As an
example, a 2D contour plot of the total valence electron density
in the plane of the molecule Si1-O-Si2 calculated using LMO's
obtained by method M in model 1 is shown in Fig.
\ref{Fig:density2D}.
 The cross-section of this plot corresponds to a solid line (the middle
cluster) in Fig. \ref{Fig:convergence}. One can see that a
considerable amount of the charge is concentrated around oxygen
atoms.

\begin{figure}
\begin{center}\includegraphics[%
  width=7cm,
  keepaspectratio]{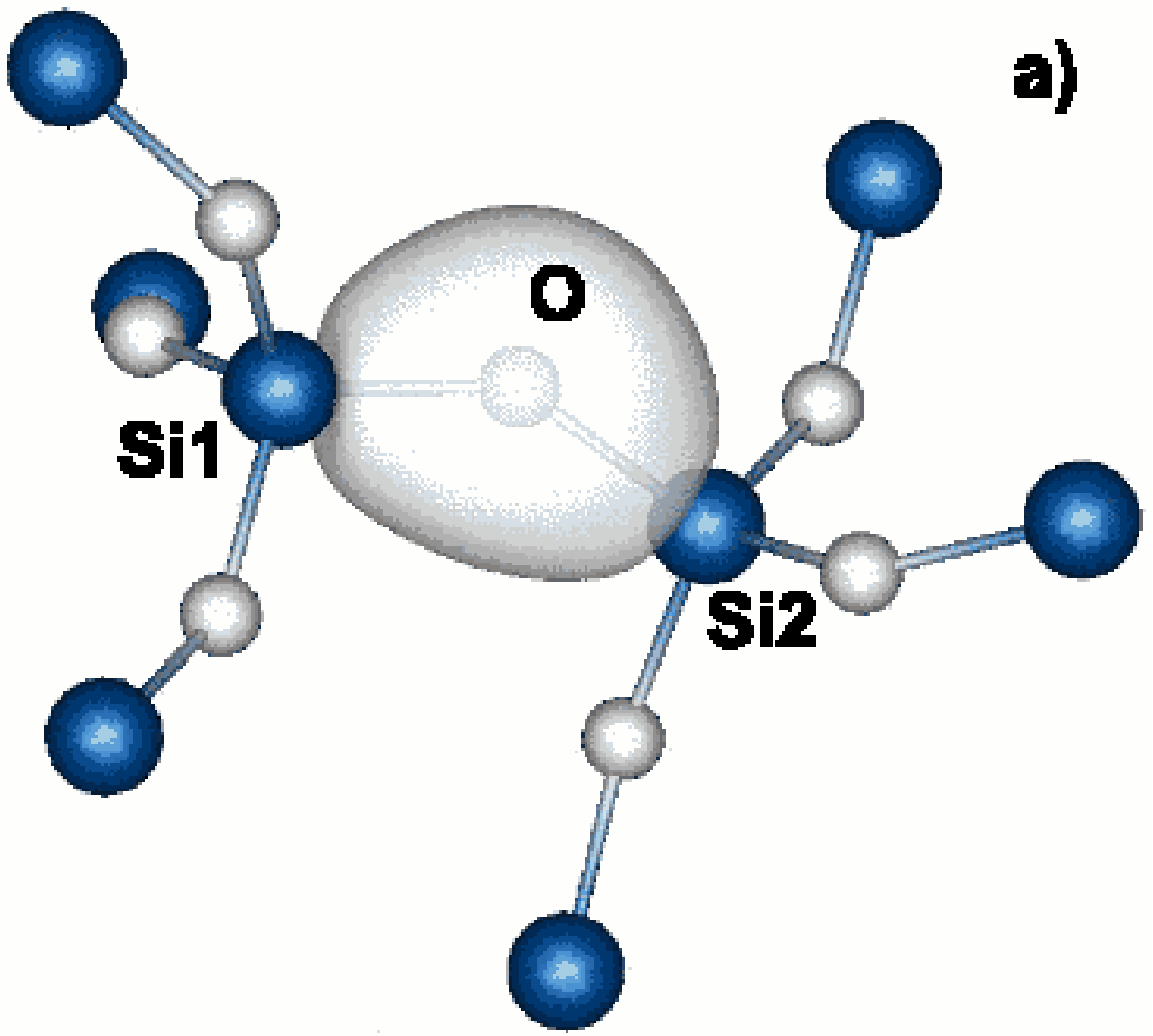}\end{center}

\begin{center}\includegraphics[%
  width=7cm,
  keepaspectratio]{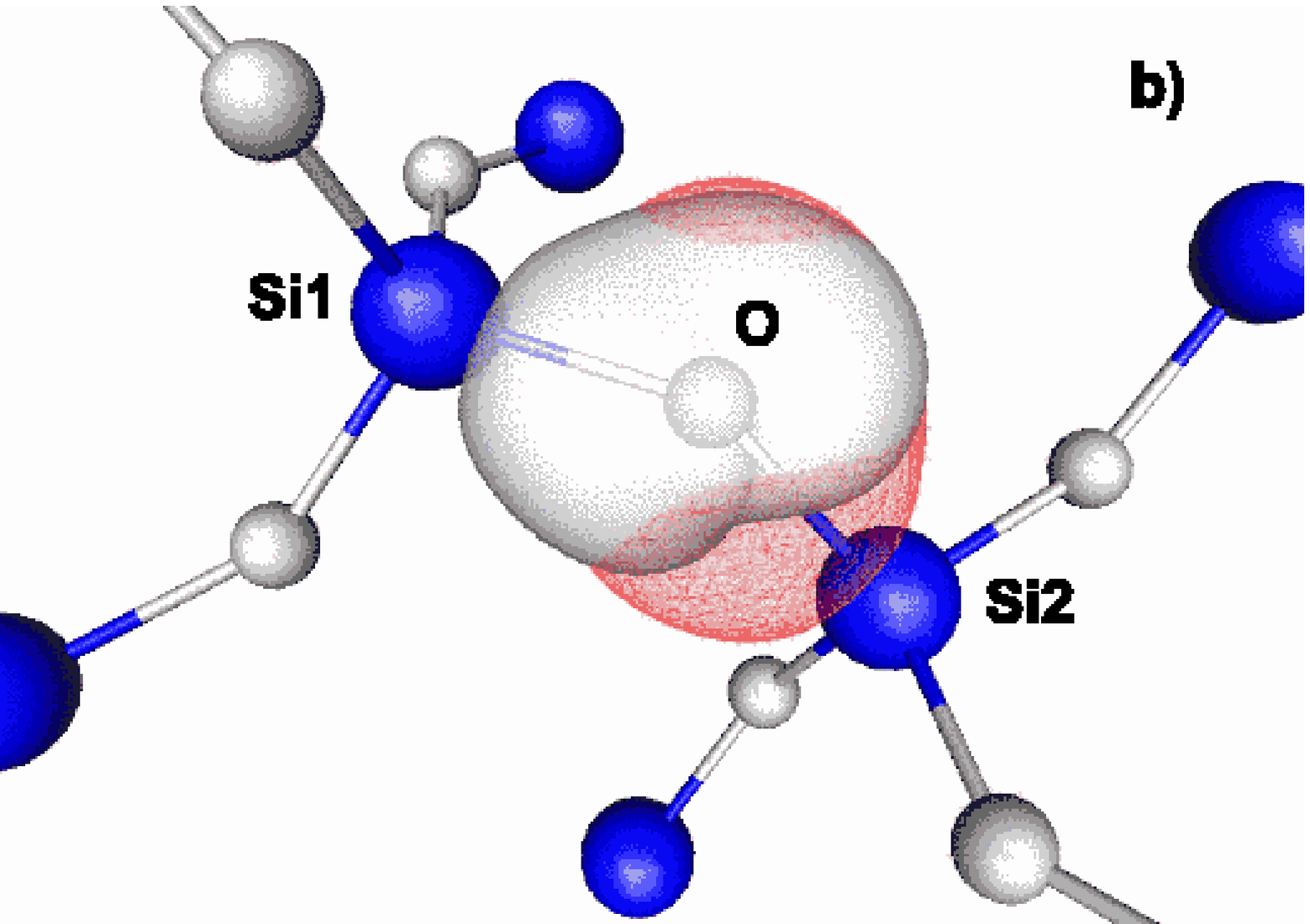}\includegraphics[%
  width=7cm,
  keepaspectratio]{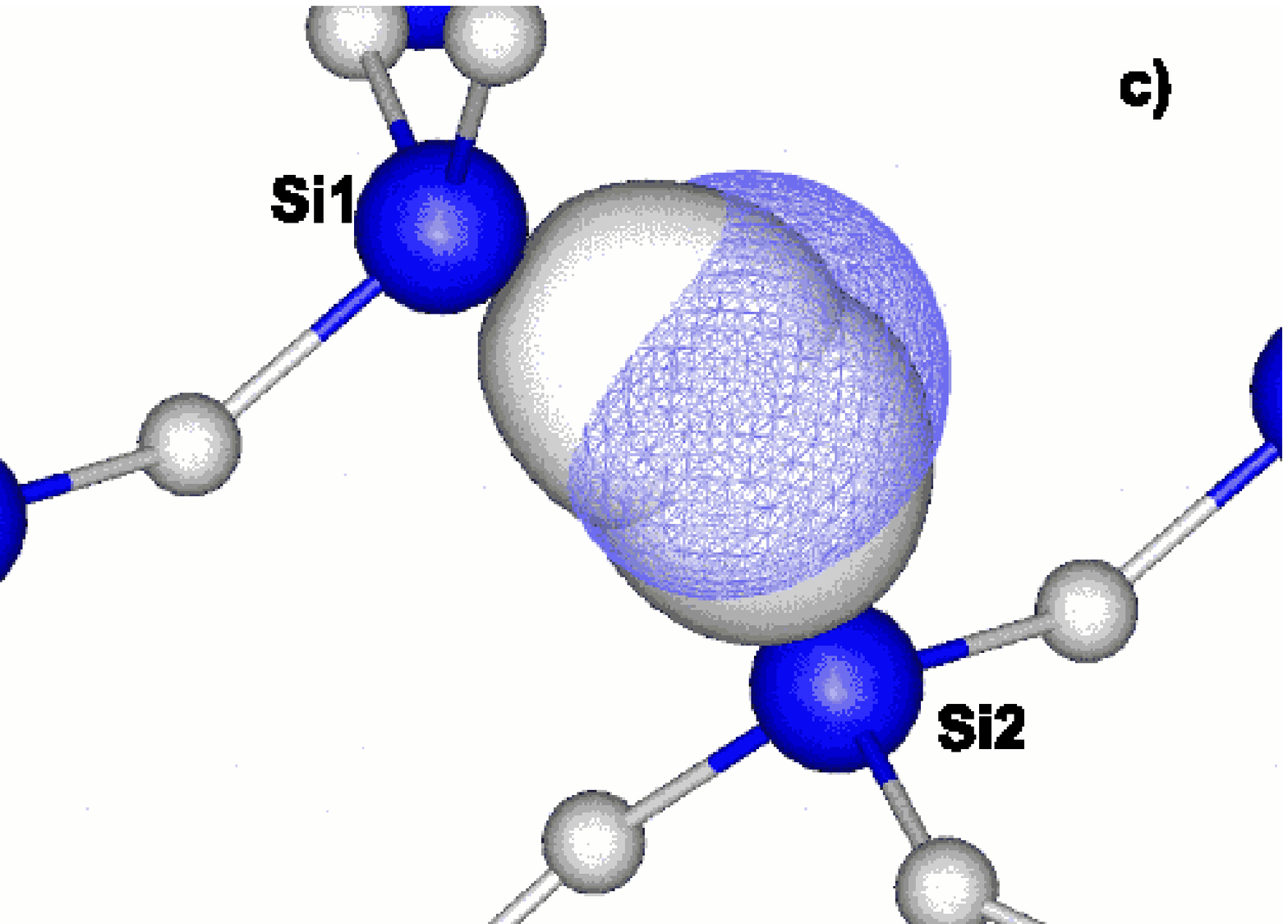}\end{center}

\caption{Constant partial density plots for three choices of
localisation regions: (a) model 1, when the whole unit
Si$_{\frac{1}{4}}$OSi$_{\frac{1}{4}}$ is associated with a single
region; (b) model 2, when two regions Si1-O and Si2-O are
identified and (c) model 3, when three regions, Si1-O, Si2-O and
O, are identified. In every case the value of the density shown is
chosen to enclose 90\% of the total electron charge associated
with the region. The localisation method M was used in each
case.\label{Fig:SE1-SE3} }
\end{figure}

The partial region densities,
$\rho_{A}(\mathbf{r})=\sigma_{A}(\mathbf{r},\mathbf{r})$ (see Eq.
(\ref{eq:region-A_density})), corresponding to each of the regions
and generated from the LMO's calculated using method M, are shown
in Fig. \ref{Fig:SE1-SE3} as closed 3D surfaces of constant
density. The value of the density is chosen in such a way so that
90\% of the region electron charge be contained inside every
surface. In the case of model 1 (the upper panel) only one density
is shown; in the case of model 2 (the middle panel) two densities
are shown simultaneously, while in the third case (model 3, the
bottom panel), all three partial densities are presented. One can
see that in the cases of models 2 and 3 the LMO's belonging to
neighbouring regions strongly overlap. As was mentioned before,
the LMO's belonging to different regions for these two models are
not orthogonal. At the same time, the overall shapes of the
density for each of the region models are very similar
demonstrating a clear aggregation around the O atom in the middle
of the Si$_{\frac{1}{4}}$OSi$_{\frac{1}{4}}$ unit in agreement
with the total density shown in Fig. \ref{Fig:density2D}.

\begin{figure}
\begin{center}\includegraphics[%
  width=100cm,
  height=7cm,
  keepaspectratio]{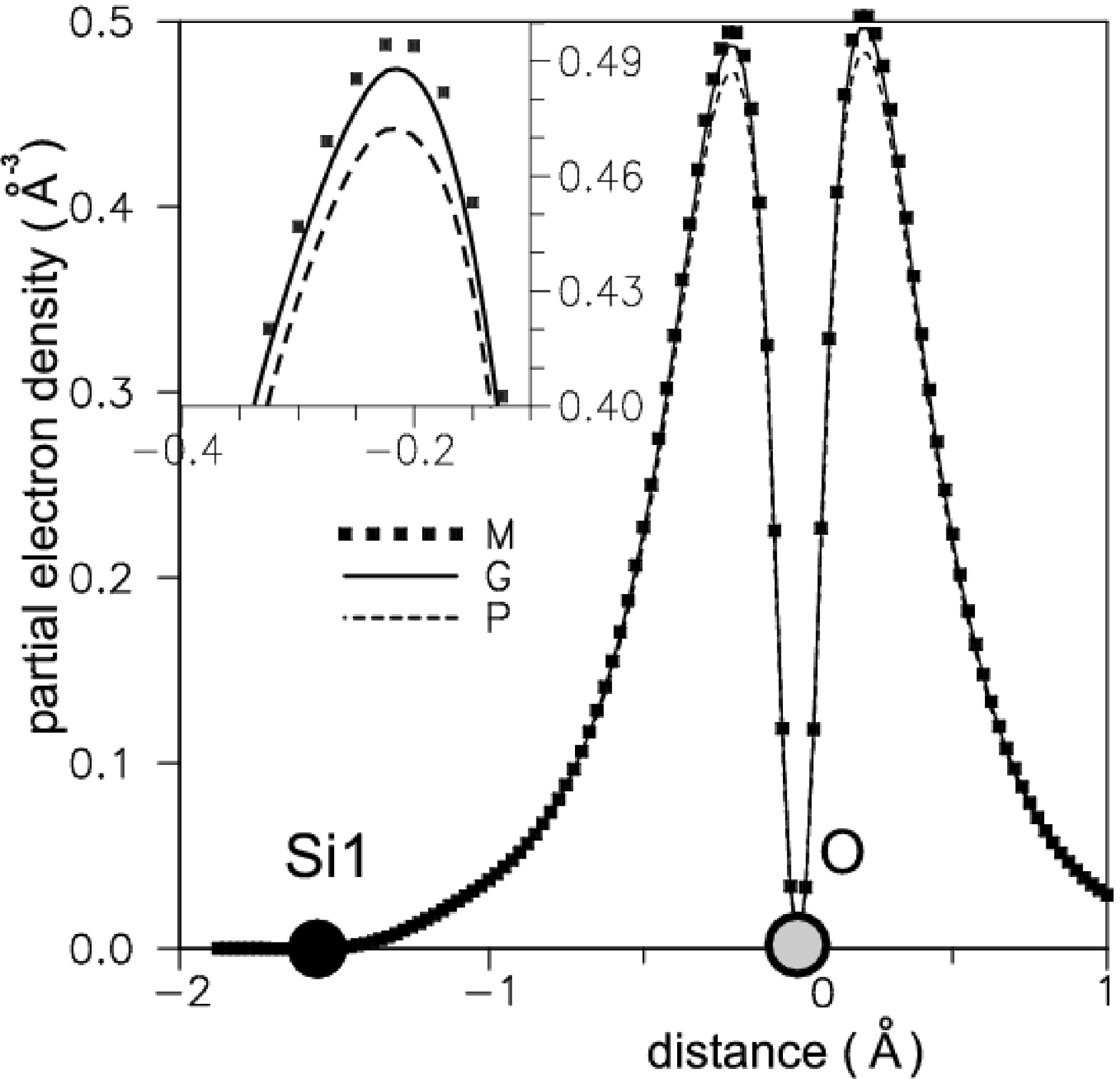}\end{center}

\caption{The partial electron density for region model 1
calculated using different localisation methods and plotted along
the Si1-O direction. \label{Fig:psi2} }
\end{figure}

A comparison of the LMO's calculated using three localisation
methods is presented in Fig. \ref{Fig:psi2} for region model 1. In
this figure, the partial densities $\rho_{A}(r)$ are shown in each
case along the Si1-O direction through the
Si$_{\frac{1}{4}}$OSi$_{\frac{1}{4}}$ unit. It is clear that the
partial density obtained from LMO's calculated using method M are
found slightly more localised, whereas the localisation obtained
using method P is slightly worse than given by the two others.
Still, the difference between the densities is extremely small so
that we can conclude at this point that all three techniques
perform practically equally well (at least for the system under
discussion).

The picture becomes more complicated, however, at least at the
first sight, when the localisation criteria of Section
\ref{sub:Localization-criteria} are applied in each of the nine
cases as shown in Table \ref{Table:localis-criteria}. Three rows
in this table correspond to the three different region models; in
each case there are four LMO's in total which are occupied by
eight electrons of the elementary
Si$_{\frac{1}{4}}$OSi$_{\frac{1}{4}}$ unit. The three columns in
the Table correspond to the three different localisation methods
used, and each of the localisation criteria is shown for every
method.

The fist criterion (the localisation index $d_{a}$) is slightly
above one in most cases, and is smaller than two in all nine
cases. This means that the LMO's are mostly localised on a single
atom with some contribution coming from the nearest atoms.
However, the maximum eigenvalues of the overlap matrix were found
to be below two only for the region model 1 and localisation
methods M and G; in the cases of models 2 and 3 eigenvalues around
three were found indicating worse localisation. Finally, the gap
between the eigenvalues of the secular problem of Eq.
(\ref{eq:eigenv-problem-matrix}), $\Delta\lambda$, was found to be
too small for the region models 2 and 3, whereas in the case of
model 1 and for all three localisation methods the gap is
considerable, especially for methods M and G. This means that the
choice of the regions in models 2 and 3 is somewhat artificial
which is not surprising because of a very strong overlap between
LMO's corresponding to the neighbouring regions in these two
models.

It follows from this analysis that the model 1, in which the whole
elementary unit Si$_{\frac{1}{4}}$OSi$_{\frac{1}{4}}$ is
considered as one region, results in the best localisation of the
LMO's, especially if methods M and G are used.

\begin{table}
\begin{center}\begin{tabular}{c|c|c||ccc|ccc|ccc}
\hline & & {\scriptsize LMO}& \multicolumn{3}{c|}{{\scriptsize
Method M}}& \multicolumn{3}{c|}{{\scriptsize Method G}}&
\multicolumn{3}{c}{{\scriptsize Method P}}\tabularnewline {\small
Model}& {\small Regions}& & {\scriptsize $d_{a}$}& {\scriptsize
$\max(\lambda(\mathbf{k}))$}& {\scriptsize $\Delta\lambda$}&
{\scriptsize $d_{a}$}& {\scriptsize $\max(\lambda(\mathbf{k}))$}&
{\scriptsize $\Delta\lambda$}& {\scriptsize $d_{a}$}& {\scriptsize
$\max(\lambda(\mathbf{k}))$}& {\scriptsize
$\Delta\lambda$}\tabularnewline \hline \hline & & {\scriptsize 1}&
{\scriptsize 1.18573}& & & {\scriptsize 1.20741}& & & {\scriptsize
1.22195}& & \tabularnewline {\scriptsize 1}& {\scriptsize
Si1-O-Si2}& {\scriptsize 2}&
 {\scriptsize 1.13868}&
{\scriptsize 1.71955}& {\scriptsize 0.92527}& {\scriptsize
1.15423}& {\scriptsize 1.76597}& {\scriptsize 0.77451}&
{\scriptsize 1.54019}& {\scriptsize 2.08836}& {\scriptsize
0.30218}\tabularnewline & & {\scriptsize 3}& {\scriptsize
1.41971}& & & {\scriptsize 1.44083}& & & {\scriptsize 1.17747}& &
\tabularnewline & & {\scriptsize 4}& {\scriptsize 1.12914}& & &
{\scriptsize 1.13788}& & & {\scriptsize 1.18254}& &
\tabularnewline & & & & & & & & & & & \tabularnewline \hline &
{\scriptsize Si1-O}& {\scriptsize 1}& {\scriptsize 1.21753}& &
{\scriptsize 0.00508}& {\scriptsize 1.20624}& & {\scriptsize
0.00223}& {\scriptsize 1.24733}& & {\scriptsize
0.00045}\tabularnewline {\scriptsize 2}& & {\scriptsize 2}&
{\scriptsize 1.15569}& {\scriptsize 2.97523}& & {\scriptsize
1.17676}& {\scriptsize 2.92854}& & {\scriptsize 1.24529}&
{\scriptsize 3.07121}& \tabularnewline & {\scriptsize Si2-O}&
{\scriptsize 1}& {\scriptsize 1.20410}& & {\scriptsize 0.01659}&
{\scriptsize 1.18245}& & {\scriptsize 0.00736}& {\scriptsize
1.24529}& & {\scriptsize 0.00108}\tabularnewline & & {\scriptsize
2}& {\scriptsize 1.16580}& & & {\scriptsize 1.19308}& & &
{\scriptsize 1.18726}& & \tabularnewline & & & & & & & & & & &
\tabularnewline \hline & {\scriptsize O}& {\scriptsize 1}&
{\scriptsize 1.07793}& & {\scriptsize 0.06952}& {\scriptsize
1.06777}& & {\scriptsize 0.04513}& {\scriptsize 1.11378}& &
{\scriptsize 0.00112}\tabularnewline {\scriptsize 3}& &
{\scriptsize 2}& {\scriptsize 1.12734}& {\scriptsize 3.43867}& &
{\scriptsize 1.11958}& {\scriptsize 3.43770}& & {\scriptsize
1.12776}& {\scriptsize 3.98629}& \tabularnewline & {\scriptsize
Si1-O}& {\scriptsize 1}& {\scriptsize 1.21753}& & {\scriptsize
0.02072}& {\scriptsize 1.20624}& & {\scriptsize 0.01363}&
{\scriptsize 1.24733}& & {\scriptsize 0.00433}\tabularnewline &
{\scriptsize Si2-O}& {\scriptsize 1}& {\scriptsize 1.20410}& &
{\scriptsize 0.00921}& {\scriptsize 1.18245}& & {\scriptsize
0.00868}& {\scriptsize 1.24529}& & {\scriptsize
0.00409}\tabularnewline & & & & & & & & & & & \tabularnewline
\hline
\end{tabular}\end{center}

\caption{Localisation criteria calculated for all region models
and localisation methods. \label{Table:localis-criteria}}
\end{table}

In spite of subtle differences in the applied localisation
criteria which seem to favour the model 1 and the methods M and G,
we stress that very good localisation of the LMO's was obtained in
all cases. This conclusion is also supported by an observation
that LMO's generated within different models (choices of the
regions) span the same Fock space. To make such a conclusion, we
calculated the projection of the LMO's of models 2 and 3 on the
space spanned by the four orthogonal LMO's obtained in model 1 and
then subtracted the projection from the original orbitals. The
calculated residual parts were found negligible in all cases.
Note, that this conclusion is not obvious because each LMO depends
on all AO's of the entire cluster.

\section{Conclusions\label{sec:Discussion-and-conclusions}}

In this paper we have calculated strongly localised molecular
orbitals (LMO's) for the SiO$_{2}$ crystal ($\alpha$-quartz) using
the method developed earlier \cite{Danyliv-LK-2004}. The starting
point for the choice of the localisation regions was an
observation that the whole crystal can be reproduced by rotating
and translating a single elementary unit
Si$_{\frac{1}{4}}$OSi$_{\frac{1}{4}}$, containing an O atom and
\emph{quarters} of the two Si atoms which the O atom is directly
connected to. Three localisation methods were applied and three
models for choosing the localisation regions were tried in each
case: (1) the whole unit was considered as one region; (2) the
unit was split into two and (3) three regions. Although in all
cases well localised orbitals were obtained, we find that the
first choice of the localisation region in which the whole unit
was chosen as a single region, is preferable.

The LMO's produced in models 2 and 3 were found very close to a
linear combination of the orthonormal LMO's obtained within model
1. If taken from the nearest regions belonging to the same
elementary unit, they appeared to have a significant overlap with
each other. On the other hand, the LMO's belonging to different
units (in either of the models) do not overlap strongly which is
confirmed by various localisation criteria applied in this work
and by the corresponding plots of the partial densities.

Since our previous calculations reported in \cite{Danyliv-LK-2004}
were done for the extreme cases of ionic (MgO) and covalent (Si)
bonding, it follows from the results of the present study that our
method is also applicable to the crystals with intermediate types
of chemical bonding.

Note that we did not consider in this study localisation method E
\cite{Danyliv-LK-2004} based on the energy minimisation of the
structure element corresponding to the chosen region. This is
because it was found in \cite{Danyliv-LK-2004} that the orbitals
obtained by this method for the Si crystal were not sufficiently
localised.

Although the LMO's reported in this work may be useful to
characterise the chemical bonding in the given crystal, they are
needed for the embedding method which is under development.
Different possibilities in choosing localisation regions open up
various ways in terminating the quantum cluster when considering,
e.g. a point defect in the crystal bulk or an adsorbed species on
the crystal surface. This variety of options may be extremely
useful in applications to keep the size of the cluster as small as
possible. If, for instance, one would like to terminate the
cluster with Si atoms, then either of the region models can be
used (model 1 would probably be more convenient as the orbitals
within each region are orthonormal). However, if a termination
with oxygens is required, then region models 2 or 3 may prove to
be more useful. In practice, a combination of terminations may be
preferable, when both Si and O atoms are used at the boundary. In
those cases all three models for choosing localisation regions may
be employed.

\subsection*{Acknowledgements}

O.D. would like to acknowledge the financial support from the
Leverhulme Trust (grant F/07134/S) which has made this work
possible. We would also like to thank S. Hao for help in the VASP
calculations.

\end{document}